\documentclass[aps,prd,twocolumn,showpacs,floatfix]{revtex4}
\usepackage{graphicx}\usepackage{amssymb,amsmath}
\usepackage{times}

\begin{document}
\title{Circumscribing Late Dark Matter Decays Model Independently}

\author{Hasan Y{\"u}ksel} 
\affiliation{Department of Physics, Ohio State University,  Columbus, Ohio 43210}
\affiliation{Center for Cosmology and Astro-Particle Physics, Ohio State University, Columbus, Ohio 43210}

\author{Matthew D. Kistler} 
\affiliation{Department of Physics, Ohio State University, Columbus, Ohio 43210}
\affiliation{Center for Cosmology and  Astro-Particle Physics, Ohio State University, Columbus, Ohio 43210}

\date{May 8, 2008}

\begin{abstract}
A number of theories, spanning a wide range of mass scales, predict dark matter candidates that have lifetimes much longer than the age of the universe, yet may produce a significant flux of gamma rays in their decays today.  We constrain such late decaying dark matter scenarios model-independently by utilizing gamma-ray line emission limits from the Galactic Center region obtained with the SPI spectrometer on INTEGRAL, and the determination of the isotropic diffuse photon background by SPI, COMPTEL and EGRET observations.  We show that no more than $\sim$5\% of the unexplained MeV background can be produced by late dark matter decays either in the Galactic halo or cosmological sources.
\end{abstract}



\maketitle

\section{Introduction}
Dark matter continues to live up to its name~\cite{Zwicky:1933gu}, despite accumulated evidence of its existence from observations of large-scale structure formation, galaxy cluster mass-to-light ratios, and galactic rotation curves.  An attractive approach towards revealing dark matter's particle identity is to search for its signature in radiation backgrounds, either from the Milky Way or in the isotropic diffuse photon background (iDPB), which can contain both cosmological and Galactic halo contributions.  Dark matter might be comprised of particles that can decay with finite lifetimes much longer than the age of the universe. In such scenarios, the resultant fluxes of decay products depend on the amount of dark matter present alone, as opposed to self-annihilation, which, being dependent on particle density squared, is very sensitive to assumptions concerning details of dark matter clustering.

A wide variety of decaying dark matter models have been examined in regards to their observable implications~\cite{decays}. Among late decaying dark matter models, sterile neutrinos with multi-keV masses have been extensively studied as dark matter candidates~\cite{sterile}, with strong constraints placed on their decays, e. g. Refs.~\cite{sterile,Watson:2006qb,Yuksel:2007xh} and references therein.  The decay of moduli dark matter~\cite{moduli} with masses of several hundred keV may contribute to the sub-MeV iDPB. The dark matter model of Ref.~\cite{Cembranos:2006gt}, inspired from minimal universal extra dimensions or supersymmetry~\cite{muedsusy}, with a mass scale of hundreds of GeV, is advocated as the source of the iDPB in the MeV range~\cite{Cembranos:2007fj}, which has yet to be accounted for with conventional sources (e.g., supernovae~\cite{snia} or active galactic nuclei~\cite{agn}) or more exotic mechanisms~\cite{exotic}.  Similarly, decaying gravitino dark matter in R-parity breaking vacua~\cite{rparity}, with multi-GeV masses, has been suggested as an explanation of iDPB spectral features in the GeV range~\cite{Ibarra:2007wg,Bertone:2007aw}.

\begin{figure}[b]
\includegraphics[width=3.25in,clip=true]{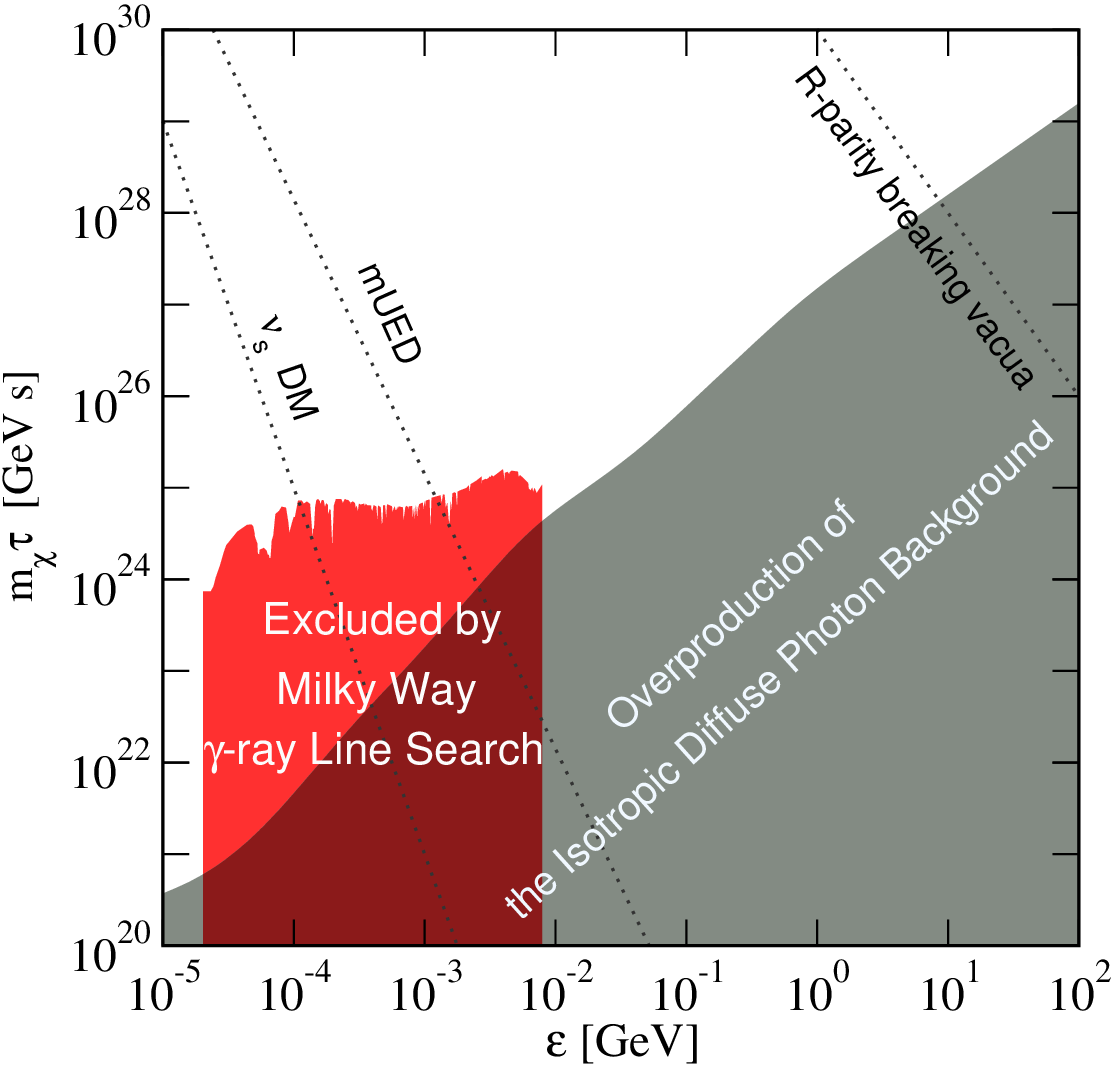}
\caption{Model-independent constraints on the product of mass and lifetime, $ m_\chi \tau$, versus the energy carried away by the monochromatic photon emission, $\varepsilon$, for a generic late-decaying dark matter model: $\chi \rightarrow \chi' + \gamma$.  Regions excluded by either the gamma-ray line emission limits from the Galactic Center region or overproduction of the isotropic diffuse photon background are shown, together with preferred ranges of parameters from three well-studied models.}
\label{fig:constraints}
\end{figure} 

Rather than focusing on a particular model, we first consider a generic decaying dark matter scenario in which the decay of the parent particle is dominated by a monochromatic photon emission. We assume that the lifetime of the parent particle, $\tau$, is much longer than the age of the universe ($\tau_0\simeq4.5 \times 10^{17}$~s), thus its cosmological abundance has not changed significantly since the time of dark matter decoupling. The decay under consideration is $\chi \rightarrow \chi' + \gamma$, where $\gamma$ is a  monochromatic photon emitted with energy $\varepsilon$. In general, $\tau$ and $\varepsilon$ will depend on the masses of the parent and the daughter particles ($m_{\chi}$, $m_{\chi'}$) and their splitting, $\Delta m = m_{\chi}-m_{\chi'}$.

The flux of photons from dark matter decays is inversely proportional to both the particle lifetime (fixing the decay rate per particle as specified by a particular theoretical model) and the mass of an individual particle (yielding the total number of particles in a fixed amount of dark matter). Thus, gamma-ray observations allow us to place constraints only on the degenerate product $m_\chi \tau$ versus $\varepsilon$, as we display in Fig.~\ref{fig:constraints}. As we will discuss in detail, below the jagged line between 0.02--8~MeV, the gamma-ray line signal from the Galactic Center (GC) region due to dark matter decays violates the corresponding limit obtained with the SPI spectrometer on INTEGRAL satellite~\cite{Teegarden}.  Additionally, the iDPB, as determined from SPI~\cite{Churazov:2006bk}, COMPTEL~\cite{Weidenspointner} and EGRET~\cite{Sreekumar:1997un,Strong:2004ry} data, is overproduced (assuming it is fully accounted by late dark matter decays in a given energy band) in the triangular region, even disregarding any contributions from known astrophysical sources.  We also show three representative scenarios, inspired by the theories of sterile neutrinos, R-parity breaking vacua, and mUED.  Since $m_{\chi}$, $m_{\chi'}$ and $\Delta m$ are not necessarily predetermined, they may be adjusted to yield the displayed curves relating $ m_\chi \tau$ and $\varepsilon$.

\begin{figure}[b]
\includegraphics[width=3.25in,clip=true]{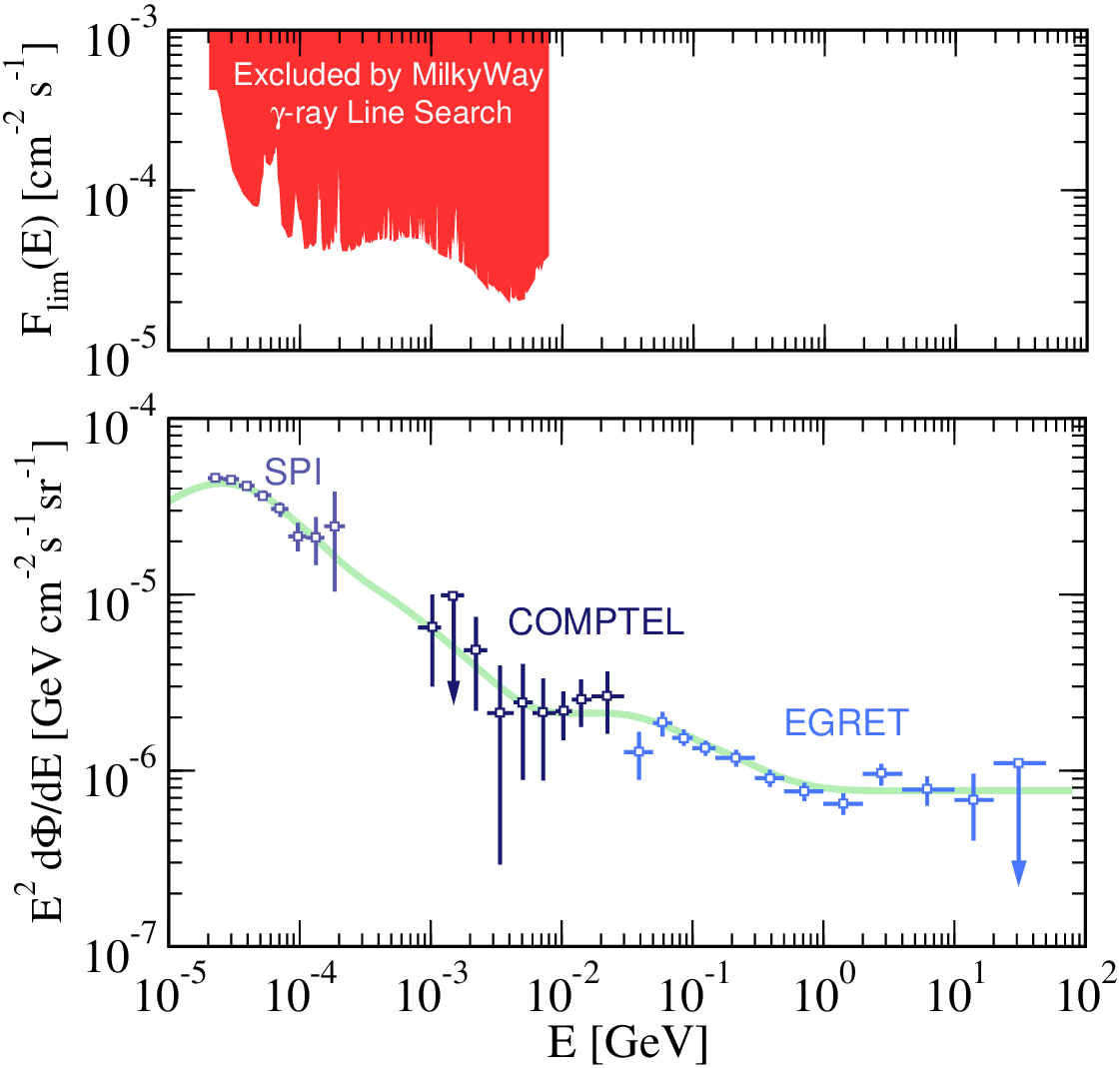}
\caption{\textbf{Top:} Limits on the diffuse gamma-ray line emission from the Galactic Center region (an angular region within a $13^{\circ}$ radius) as adopted from Ref.~\cite{Teegarden}. \textbf{Bottom:} Representative measurements of the diffuse photon background from SPI~\cite{Churazov:2006bk}, COMPTEL~\cite{Weidenspointner} and EGRET~\cite{Strong:2004ry} in the energy range around 0.01~MeV--100~GeV. The thick solid line, summarizing the overall trend of the data, is to be compared to predictions of decaying dark matter scenarios.
}
\label{fig:dpb}
\end{figure} 

\section{Milky Way Gamma-Ray Line Search}
A monochromatic line will be most detectable locally, where cosmological redshifting is of no concern.  Fortunately, a search for diffuse gamma-ray line emission in the energy range 0.02-8~MeV from the GC region has been conducted by Teegarden and Watanabe using the SPI spectrometer on the INTEGRAL satellite~\cite{Teegarden}, which recovered the known astrophysical diffuse line fluxes, such as the 511~keV positron annihilation line~\cite{511line}. The excellent energy resolution of SPI enabled them to place very strict constraints on potential unidentified emission lines, with an energy dependent 3.5~$\sigma$ flux limit, ${\cal F}_{lim}(E)$, from an angular region within a $13^{\circ}$ radius of the GC (which we refer as the GC region).  This limit, reproduced in the top panel of Fig.~\ref{fig:dpb}, can be compared to the expected gamma-ray flux arising from late dark matter decays in the GC region, which we calculate following the methods in Ref.~\cite{Yuksel:2007xh}.

We first define a dimensionless line-of-sight integral at an angle $\psi$ relative to the GC,
\begin{equation}
{\cal J}(\psi) = \frac{1}{\rho_{sc} R_{sc}} \int_{0}^{\ell_{max}} d\ell \; 
\rho\left(\sqrt{R_{sc}^2-2\, \ell\, R_{sc}\cos\psi+\ell^2} \right)\,,
\label{eq:Jintegral}
\end{equation}
where $\rho$ is the density of the dark matter in the halo as a function of the distance from the GC.  This is normalized to the dark matter density ($\rho_{sc} = 0.3$ GeV cm$^{-3}$) at the solar circle ($R_{sc} = 8.5$~kpc) so that $\rho_{sc} R_{sc} \simeq 8\times 10^{21}$ GeV cm$^{-2}$.  Note that this arbitrary normalization is needed to make ${\cal J}$ dimensionless and will be canceled-out later.  The upper limit of this integration,
\begin{equation}
\ell_{max}= \sqrt{(R_{MW}^2-\sin^2 \psi R_{sc}^2)} + R_{sc} \cos \psi \,,
\label{eq:ellmax}
\end{equation}
depends on $R_{MW}$, the assumed size of the halo. ${\cal J}$ is relatively insensitive to $\ell_{max}$ as long as $R_{MW}$ is large.  The intensity of photons (number flux per solid angle) from the same direction,
\begin{equation}
{\cal I}(\psi) = \frac{\rho_{sc} R_{sc}} {4\pi  m_\chi \tau} {\cal J}(\psi)\,,
\label{eq:intensity}
\end{equation}
can be integrated over a circle of radius $\psi$ around the GC (covering a patch of area $\Delta \Omega = 2 \pi (1 - \cos \psi)$) to obtain the corresponding total flux,
\begin{equation}
{\cal F}=  \int_{\Delta \Omega}  d\Omega' \; {\cal I}(\psi')
= \frac{\rho_{sc} R_{sc}} {4\pi m_\chi \tau} \int_{\Delta \Omega}  d\Omega' \; {\cal J}(\psi')
\label{eq:decflux}
\end{equation}

The limit reported in Ref.~\cite{Teegarden} has been obtained by subtracting the average flux measured at regions away from the GC region ($\psi > 30^\circ$) from the average flux measured inside the GC region ($\psi < 13^\circ$) to eliminate instrumental backgrounds.  Thus, the constraining power of this limit for decaying dark matter scenarios depends on the enhancement of the expected signal towards the GC region.  Both theoretical and observational studies strongly suggest that the central regions of dark matter halos are significantly denser and, moreover, the column depth is higher towards the GC direction relative to off-axis lines-of-sight.  We have reproduced the impact of this subtraction (see Ref.~\cite{Yuksel:2007xh} for details) by calculating a parameter 
\begin{equation}
\zeta_{\rm lim}= \int_{\Delta\Omega} d\Omega' \, [{\cal J}(\psi') - {\cal \overline J}_{>30^{\circ}}] \,.
\label{eq:deltaF}
\end{equation}
which ranges between $\sim0.5-1.5$ for various dark matter halo fitting profiles commonly used in the literature~\cite{haloprofiles}. Here ${\cal \overline J}_{>30^{\circ}}$ is the average of ${\cal J}$ away from the GC region. We also note that the results that we adopted from Ref.~\cite{Teegarden} are based on an assumption that the expected line signal has a Gaussian source profile, while a flat source profile could yield limits that are weaker by up to a factor of $\sim$2 (e.g., see Fig.~12 of Ref.~\cite{Boyarsky:2007ge}).  Moreover, one would expect to see these limits improve as the amount of available data increases in time~\cite{Boyarsky:2007ge}.  In the rest of our study, we choose a conservative value, $\zeta_{\rm lim}\simeq 0.5$, which can be realized only for profiles that are rather flat inside the solar circle.   While this mostly protects our conclusions from uncertainties in the halo profile, our subsequent result can be easily rescaled for a different value.

The predicted gamma-ray emission line flux due to dark matter decays at a given $\varepsilon$ must not exceed the corresponding limits from the GC region, thus 
\begin{equation}
\frac{\rho_{sc} R_{sc}} {4\pi m_\chi \tau}\, \zeta_{\rm lim} < {\cal F}_{\rm lim}(E=\varepsilon) \,.
\label{eq:gclimit1}
\end{equation}
Rearranging this equation yields our model independent constraint,
\begin{equation}
{ m_\chi \tau} > \frac{\rho_{sc} R_{sc} \zeta_{\rm lim}}{ 4\pi {\cal F}_{\rm lim}( \varepsilon )} 
\simeq \frac{3\times 10^{20}\, {\rm GeV \, cm}^{-2}}{{\cal F}_{\rm lim}( \varepsilon )} \,,
\label{eq:gclimit2}
\end{equation}
as shown in Fig.~\ref{fig:constraints} (region below the jagged line). The expected dark matter decay flux is inversely proportional to $m_\chi \tau$, which leads to an overproduction of gamma rays for $ m_\chi \tau \lesssim 10^{25}$~GeV~s in the energy range 0.02-8~MeV. Thus the area below the jagged line is excluded by the the diffuse gamma-ray line emission limits from the GC region.

\section{Isotropic Diffuse Photon Background}
While stringent limits on line emission from the GC region are only available in a rather limited energy range (0.02--8~MeV), the iDPB is measured over a broad range by many instruments.  In the bottom panel of Fig.~\ref{fig:dpb}, we display three recent determinations of iDPB in different ranges of energy from SPI~\cite{Churazov:2006bk}, COMPTEL~\cite{Weidenspointner} and EGRET~\cite{Strong:2004ry}, which are consistent with others measurements (see, e.g., Ref.~\cite{Gruber:1999yr,Strong:2005zn}). The thick dotted line represents the global trend of the data to be used for comparison. We choose the terminology ``isotropic diffuse'' photon background (iDPB), as opposed to ``cosmic'' or ``extragalactic'', since the contribution from sources in the Milky Way or its halo is not clear, and iDPB can include gamma-ray line signals that could not have been resolved by COMPTEL or EGRET.  While it is generally thought that AGN are responsible for the emission in the $\sim$~keV~\cite{lowe} and $\sim$~GeV~\cite{highe} ranges, the origin of the iDPB, especially in the MeV regime, is far from being settled, with various scenarios having been entertained~\cite{snia,agn,exotic}.  It is then of interest to determine just how much of the iDPB can possibly be accounted for by late decaying dark matter.

\subsection{Dark Matter Decays in the Halo}
While the photon signal from dark matter decays in the Galactic halo is enhanced towards the GC, as has been utilized for our constraints in the earlier section, it also contains an apparently isotropic contribution.  The limited energy resolution of past gamma-ray detectors could not distinguish monochromatic line emission from the Galactic halo from a truly cosmological signal.  The intensity of the isotropic halo contribution, ${\cal I}_{\rm iso}$, can be estimated from a line of sight integration in the anti-GC direction, ${\cal J}_{\rm iso}={\cal J}(180^{\circ})\sim 1$, as this is the minimum contribution from the dark matter halo of the Milky Way.  Regardless of the underlying halo profile, this number is relatively robust, being mostly dependent on the dark matter density at the solar circle.  The intensity of this isotropic component is
\begin{equation}
{\cal I}_{\rm iso} = \frac{\rho_{sc} R_{sc}} {4\pi  m_\chi \tau} {\cal J}_{\rm iso}\,.
\label{eq:intensity2}
\end{equation}

\begin{figure}[b]
\includegraphics[width=3.25in,clip=true]{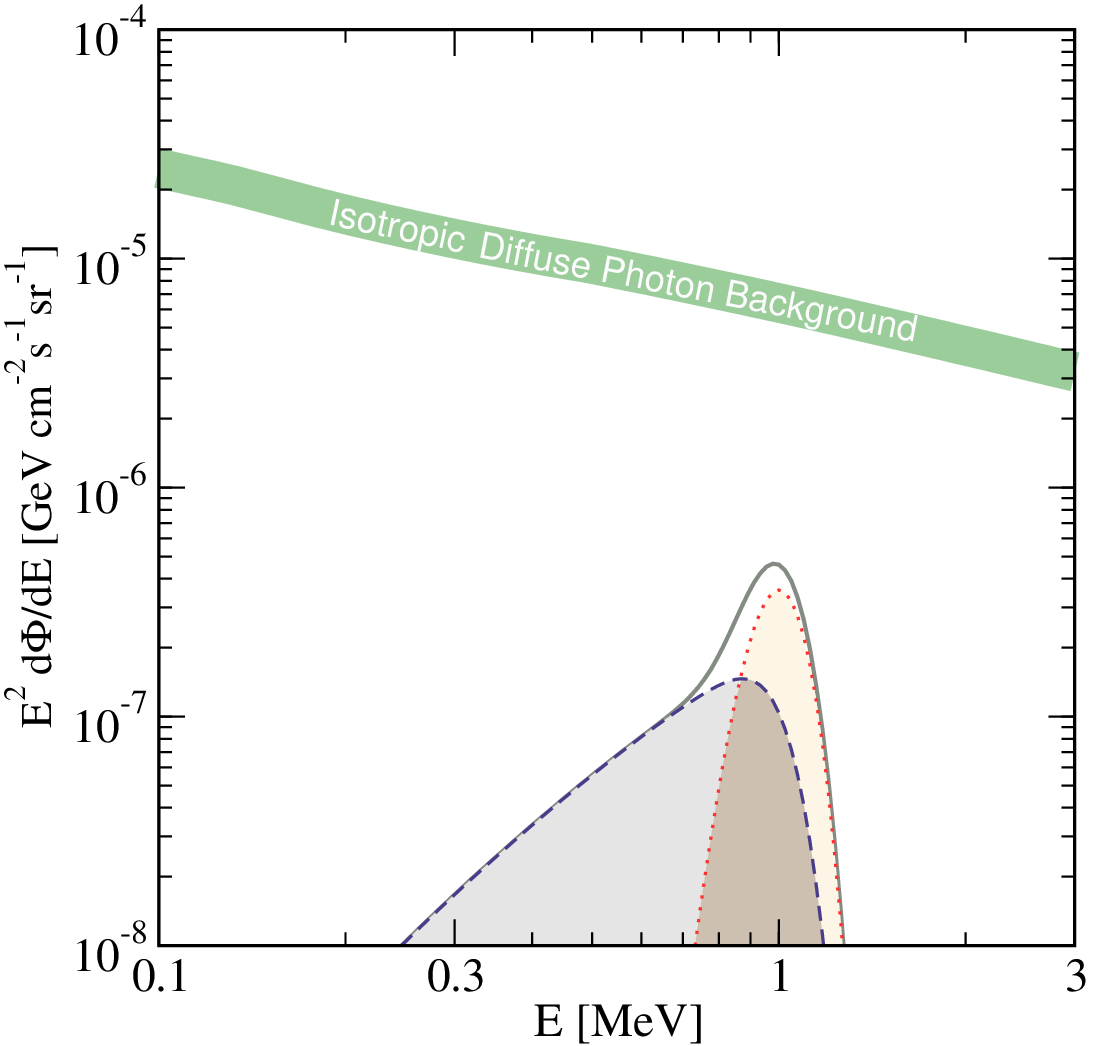}
\caption{Photon spectrum from isotropic Galactic halo decays (dotted line) for $\varepsilon=1$~MeV, with $ m_\chi \tau \simeq 7 \times 10^{24}$~GeV~s chosen from Fig.~\ref{fig:constraints} such that the line emission bounds from the Galactic Center region are saturated.  Also displayed are the spectra from cosmological decays (dashed line) and the total spectrum (solid line), which falls well below the isotropic diffuse photon background (thick solid line).
\label{fig:spectrum} }
\end{figure}

We present a representative spectrum for this isotropic signal in Fig.~\ref{fig:spectrum} (dotted line), after convolution with a Gaussian of $\sim10$\% width to simulate the energy resolution of a typical detector.  We have chosen $\varepsilon =1$~MeV, with $m_\chi \tau = 7 \times 10^{24}$~GeV~s, the maximum value allowed by the the line emission bounds from the GC region (Fig.~\ref{fig:constraints}).  For these parameters, the isotropic contribution of the dark matter decays in the Galactic halo alone to the iDPB is less than 2\% (in a bin of logarithmic width 0.4 dex centered around $\varepsilon =1$~MeV). Note that the average flux expected from the decays in the Galactic halo (which is more directional, peaking toward the GC region) can be at most several times larger than this isotropic component since we are dealing with decaying dark matter particles (contrary to self-annihilating dark matter, which is highly sensitive to the details of dark matter clustering).

\subsection{Cosmological Dark Matter Decays}
We now evaluate the contribution of truly cosmological dark matter decays to the iDPB.  For late decaying particles ($\tau \gg \tau_0$), the comoving dark matter density has remained nearly constant since the early universe.  The comoving decay rate is then simply proportional to the dark matter fraction ($\Omega_{\chi}\simeq 0.25$) of the critical density of the universe, $\rho_c$, and is given as $\rho_c \Omega_{\chi}/ (m_\chi \tau)$.  The diffuse gamma-ray flux (per solid angle per unit energy) arising from the decays  can be calculated by considering the contributions from all redshifts (analogous to~\cite{cosmic}),
\begin{eqnarray}
\frac{d\Phi}{dE} &=& \frac{1}{4\pi}\frac{c}{H_0} \int \frac{dz}{h(z)} \frac{\rho_c \Omega_{\chi}}{m_\chi \tau} \, \delta(E(1+z)-\varepsilon) 
\,,
\end{eqnarray}
where $h(z)= [(1+z)^3 \Omega_{M} + \Omega_\Lambda]^{1/2}$, $\Omega_{M}\simeq0.3$, $\Omega_{\Lambda} \simeq 0.7$, $H_0=70\, {\rm\,km \,s}^{-1} {\rm Mpc}^{-1}$, and $c=3\times 10^{10} \; {\rm cm \,s^{-1}}$ (so that $c/H_0 \simeq 1.3 \times 10^{28} {\rm cm}$ and $\rho_c=5.3 \times 10^{-6}$ GeV cm$^{-3}$).  The integration can be eliminated after using the $\delta$-function identity; $\delta(ax-b)=\delta(x-b/a)/a$, simplifying the result to
\begin{equation}
\frac{d\Phi}{dE} =\frac{1}{4\pi}\frac{c}{H_0} \frac{\rho_c \Omega_{\chi}}{m_\chi \tau}
\frac{1}{E} \frac{\Theta(\varepsilon - E)}{\sqrt{(\varepsilon/E)^3 \Omega_{M} + \Omega_\Lambda} } \, ,
\end{equation}
where $h$ is substituted and $\Theta$ is a step function.  We show this, using the same parameters as in the preceding subsection and again smoothing with a $\sim10$\% Gaussian, in Fig.~\ref{fig:spectrum} (dashed line).  As seen in the figure, this cosmological flux is slightly lower than the isotropic contribution from the Galactic halo, and their sum (solid line) still falls well short of the observed signal, restricting their combined contribution to the iDPB to be less than 4\%.

To quantify and generalize our observations, we calculate the expected total (cosmological plus isotropic Galactic halo) spectrum for all values of $ m_\chi \tau$ and compare to the iDPB (as denoted by the thick trend curve in Fig.~\ref{fig:dpb}), integrating both in a bin of logarithmic width 0.4 dex centered around $\varepsilon$.  This choice encompasses most of the expected signal where the decay spectrum peaks, and both exceeds the experimental energy resolution and the uncertainties on the determination of the iDPB. In Fig.~\ref{fig:constraints}, the region in which dark matter decays overproduce the iDPB is shown (triangular region).
  
Above this region, decaying dark matter alone cannot fully account for the iDPB.  In fact, since there should be additional contributions from AGN at both low and high energies~\cite{lowe,highe}, the actual bound on the parameter $ m_\chi \tau$ will be even more stringent than the one presented.  Combining the iDPB overproduction constraint and the gamma-ray line emission limit from the GC region  model-independently excludes a sizable region in the parameter space of $ m_\chi \tau$ versus $\varepsilon$, with the latter picking up when the former is exhausted at $\varepsilon \simeq 8$~MeV.

\section{Decaying Dark Matter Models}
While we derive our constraints for a decay scenario that is dominated by monochromatic photon emission, there may be additional modes of decay or self-annihilations producing other signals. Our constraints on the lifetime of the dark matter candidate via monochromatic photon emission could be generalized to the total lifetime including other decay channels, as long as the latter is long enough to justify the assumption that the cosmological abundance of the parent particle has not changed significantly.

For the generic decay we are considering, the energy of the emitted photon is dictated by the splitting, $\Delta m$, as follows.  When $\Delta m \ll m_{\chi'}$ (or equivalently  $m_{\chi} \simeq m_{\chi'}$), the recoil of the daughter can be neglected, so that $\varepsilon \rightarrow \Delta m$.  For $\Delta m \gg  m_{\chi'}$ (or  $m_{\chi} \gg m_{\chi'}$), two relativistic particles are produced, so that $\varepsilon \rightarrow \Delta m/2 \simeq m_{\chi}/2 $.  Generally, models lie in one of these two regimes.  To emphasize the generality of our constraints, now we discuss particular scenarios.

\begin{figure}[t]
\includegraphics[width=3.25in,clip=true]{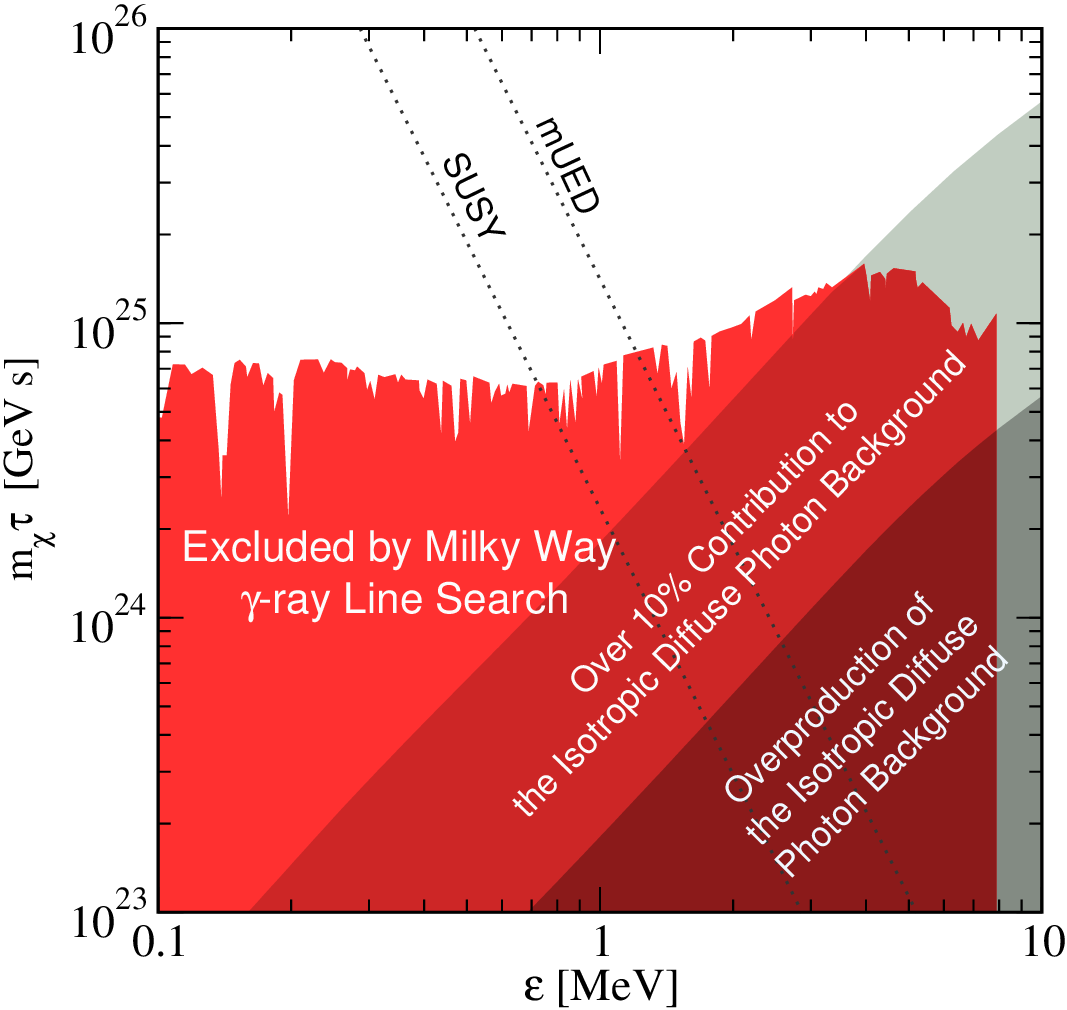}
\caption{Similar to Fig.~\ref{fig:constraints}, focusing upon the MeV range of $\varepsilon$.  The contribution of late dark matter decays to the isotropic diffuse photon background is 10\% or more in the diagonal band.  SUSY and mUED inspired decaying dark matter models of Ref.~\cite{Cembranos:2007fj} cannot make significant contribution to the iDPB while abiding by the gamma-ray line emission limits from the Galactic Center region.}
\label{fig:parameters}
\end{figure}

For example, WIMPs with weak-scale masses and cross sections may have monochromatic decays.  The decay process between the two lightest Kaluza-Klein (KK) particles, the KK hypercharge gauge boson, $B^1$ and KK graviton, $G^1$ in mUED models, and the decay between the two lightest particles in SUSY theories, the Bino-like neutralino, $\tilde{B}$ and gravitino, $\tilde{G}$ are well-studied examples.  The mass scales of these candidates are $\sim$800~GeV for the former and $\sim$80~GeV for the latter.  The decay rates in these theories~\cite{Cembranos:2007fj} are highly suppressed due to the weakness of gravity and is given by
\begin{equation}
\tau \simeq  \frac{4.7 \times 10^{22}~{\rm s}}{b} \left(\frac{\Delta m}{\rm MeV} \right)^{-3} \, .
\label{eq:tauwimp}
\end{equation}
where the parameter $b$ is identified as (2, 10/3, 1, 2) for each of the decay reactions ($G^1 \rightarrow B^1 +\gamma$, $B^1 \rightarrow G^1 + \gamma$, $\tilde{G} \rightarrow \tilde{B} + \gamma$, $\tilde{B} \rightarrow \tilde{G} +\gamma$).  The lifetime requirement of $\tau \gg \tau_0$ translates into  $\Delta m < 30$ MeV. Since $\Delta m \ll m_\chi$, the energy carried away by the emitted photon is $\varepsilon \simeq \Delta m$. Eq.~(\ref{eq:tauwimp}) can be rearranged as
\begin{equation}
m_\chi \,\tau \simeq  {4.7 \times 10^{22}~{\rm s}} \left(\frac{m_\chi}{b}\right) \left(\frac{\varepsilon} {\rm MeV}\right)^{-3}\,,
\end{equation}
which relates $m_\chi \tau$ to $\varepsilon$ in terms of a single parameter: ${m_\chi}/{b}$. We plot $m_\chi \tau$ versus $\varepsilon$ in Fig.~\ref{fig:parameters} for ${m_\chi}/{b} \simeq 300$~GeV to represent mUED. One sees that the Milky Way constraint requires $\varepsilon \leq 1.5$~MeV, which is a very strict limit as the lifetime is proportional to $\varepsilon^{-3}$, i.e., the decay rate increases by almost an order of magnitude from 1~MeV to 2~MeV. This translates to the restriction of $\Delta m \lesssim 1.5$~MeV, which is far stricter than the necessary condition to have a long-lived candidate, $\Delta m < 30$~MeV.

In the R-parity violating supersymmetric extension of the standard model, the lightest supersymmetric particle is again a gravitino that might not be stable on cosmological time scales against decay into a photon and neutrino ($\tilde{G} \rightarrow \nu + \gamma$) through a small photino-neutrino mixing $|U_{\tilde\gamma\nu}|$.  The lifetime of the gravitino in this model~\cite{Bertone:2007aw} is 
\begin{eqnarray}
\tau\simeq  3.8\times 10^{27} \, {\rm s} 
\left(\frac{|U_{\tilde\gamma\nu}|}{10^{-8}}\right)^{-2} \left(\frac{m_{\chi}}{10 \mbox{ GeV}} \right)^{-3}\,,
\end{eqnarray}
with the resulting photon and neutrino each carrying an energy of $\varepsilon = m_{\chi}/2$.  We can rewrite this equation in terms of $m_\chi \tau$ versus $\varepsilon$ as
\begin{equation}
m_\chi \tau \simeq 10^{14}~{\rm GeV \, s} \, \left(|U_{\tilde\gamma\nu}|\right)^{-2} \left(\frac{\varepsilon } {\rm GeV} \right)^{-2} \,.
\label{eqm}
\end{equation}

We plot this relation for $|U_{\tilde\gamma\nu}|=10^{-8}$ in Fig.~\ref{fig:constraints}, which shows that the contribution of this decay model to the iDPB will be significant around $\varepsilon\sim 5$~GeV (corresponding to $m_\chi\sim 10$~GeV) agreeing with Ref.~\cite{Bertone:2007aw}.  Slightly above/below $m_\chi\sim 10$~GeV, either its contribution is negligible or vastly overproduces the iDPB.

Dark matter models involving keV-mass sterile neutrinos, in their simplest description, require only two parameters, the sterile neutrino's mass and mixing with active neutrinos.  The decay chain for sterile neutrinos is $\nu_s \rightarrow \nu_{\rm e,\mu ,\tau} + \gamma$, with a radiative lifetime~\cite{lifetime} (for Dirac neutrinos)  of 
\begin{equation}
{\tau}= 1.5 \times 10^{22} {\rm\ s}
\left( \sin^{-2} 2\theta \right) \left(\frac{m_s}{\rm keV}\right)^{-5} \,.
\end{equation}
This can similarly be rearranged, keeping in mind that the energy of the parent sterile neutrino is split equally between the photon and the daughter neutrino ($\varepsilon = m_s/2$), as
\begin{equation}
m_s \tau \simeq  10^{15}~{\rm GeV \, s}\,\left( \sin^{-2} 2\theta \right) \, \left(\frac{\varepsilon}{\rm keV}\right)^{-4} \,,
\label{eq:tsnu}
\end{equation}
which has only one free parameter, ${\sin^2 2\theta}$. For illustration, we plot Eq.~(\ref{eq:tsnu}) in Fig.~\ref{fig:constraints} for ${\sin^2 2\theta}=10^{-18}$.  As seen in the figure, and has been established in Ref.~\cite{Yuksel:2007xh,Boyarsky:2007ge}, the gamma-ray line emission limit from the Galactic Center region provides quite stringent restrictions on sterile neutrino dark matter, which can be several orders of magnitude stronger than constraints from overproduction of the iDPB. Interestingly, all three models we have discussed have the form $m_\chi \tau \propto \varepsilon^{-\alpha} $, where $\alpha=$~3,~2,~4 respectively, as can also be noticed from the varying slopes of the lines representing the models in Fig.~\ref{fig:constraints}.

\section{Conclusions}
Predictions of photon fluxes from dark matter decays are considerably more robust than those from annihilations, due to a lesser dependence upon theoretical uncertainties in the distribution and clustering of dark matter.  We have shown that the gamma-ray line emission limits from the Galactic Center region, along the isotropic diffuse photon background, allow for stringent constraints to be placed on late decaying dark matter scenarios that produce monoenergetic photons. We emphasize that the Galactic and cosmic constraints are not independent of each other, with the GC region providing stronger limits in its range of applicability due to new spectroscopic data. Rather than attempting to explain various gamma-ray phenomena with a specific model, we report model-independent bounds on the decaying dark matter parameter space (as defined by $ m_\chi \tau$ versus $\varepsilon$).  Our general constraints are applicable to a number of models, and can be used as a guide for future model building.  Upcoming gamma-ray telescopes with improved energy resolution, such as GLAST~\cite{Gehrels:1999ri} or ACT~\cite{Boggs:2006mh}, can improve upon these bounds, particularly by making use of the unique spectral shape and directionality of decays from the Galactic halo.

One interesting application of our study is to assess the recent suggestion that cosmological late dark matter decays can explain the isotropic diffuse photon background in the 1-5~MeV range, whose origin remains a mystery~\cite{Cembranos:2007fj}.  We plot $m_\chi \tau$ versus $\varepsilon$ in Fig.~\ref{fig:parameters} for ${m_\chi}/{b} \simeq 300$~GeV  and ${m_\chi}/{b} \simeq 50$~GeV to represent the aforementioned mUED and SUSY models of Ref.~\cite{Cembranos:2007fj}, respectively.  We also show the range of parameters, $m_\chi \tau$ versus $\varepsilon$, that can lead to a substantial ($>10$\%) contribution to the iDPB (shaded diagonal band) or overproduce them (triangular region) through the sum of the local decays (Galactic halo) or decays from truly cosmological sources (all distant dark matter halos). The region excluded by the gamma-ray line emission limits from the GC region is below the jagged line. As seen in the figure, even the combined emission from the Galactic halo and cosmological sources due to either the mUED or SUSY inspired decaying dark matter models of Ref.~\cite{Cembranos:2007fj} cannot make a significant contribution to the iDPB while abiding by the gamma-ray line emission limits from the GC region.  The mUED model can contribute to the iDPB only for $\Delta m \lesssim 1.5$~MeV with a contribution of $\lesssim $ 5\%, while the SUSY model is even more severly constrained.  Even relaxing our assumptions on the distribution of dark matter in the halo does not increase these fractions dramatically, thus, dark matter cannot decay in the late universe at a high enough rate to make a prominent contribution to the iDPB in the MeV range.

We thank Shmuel Nussinov, Gary Steigman, Louie Strigari, Ak{\i}n Wingerter and especially John Beacom for fruitful discussions. HY is supported by National Science Foundation under CAREER Grant PHY-0547102 to JB; MDK is supported by the Department of Energy grant DE-FG02-91ER40690; both by CCAPP at OSU.


\end{document}